# Zero-inertia Offshore Grids: N-1 Security and Active Power Sharing

Georgios S. Misyris, *Student Member, IEEE*, Andrea Tosatto, *Student Member, IEEE*, Spyros Chatzivasileiadis, *Senior Member, IEEE,* and Tilman Weckesser, *Member, IEEE*

*Abstract*—With Denmark dedicated to maintaining its leading position in the integration of massive shares of wind energy, the construction of new offshore energy islands has been recently approved by the Danish government. These new islands will be zero-inertia systems, meaning that no synchronous generation will be installed in the island and that power imbalances will be shared only among converters. To this end, this paper proposes a methodology to calculate and update the frequency droops gains of the offshore converters in compliance with the N-1 security criterion in case of converter outage. The frequency droop gains are calculated solving an optimization problem which takes into consideration the power limitations of the converters as well as the stability of the system. As a consequence, the proposed controller ensures safe operation of off-shore systems in the event of any power imbalance and allows for greater loadability at pre-fault state, as confirmed by the simulation results.

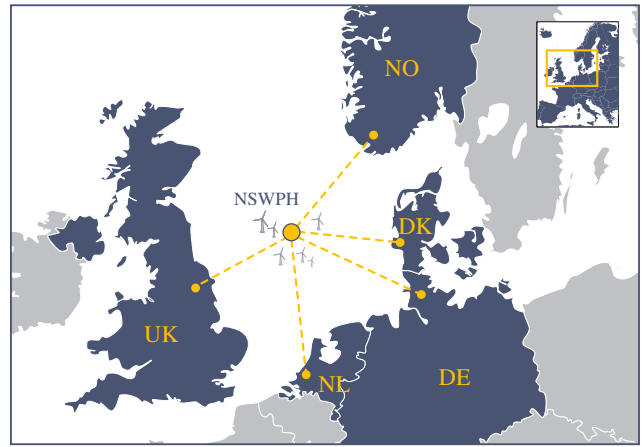

Fig. 1. North Sea Wind Power Hub concept.

## I. Introduction

SINCE the 1990s, Denmark has taken a leading role in the wind energy industry, in particular in the offshore sector. With the new Global Climate Action Strategy aiming at mitigating climate change, investments on Renewable Energy Sources (RES) will further increase. As part of this strategy, the Danish Government has approved the plans for the world's first two energy islands: Bornholm and the North Sea Wind Power Hub (NSWPH). The latter is the result of the international collaboration between the Dutch, German and Danish Transmission System Operators (TSOs) with the aim of installing more than 30 GW of wind power in the North Sea [1]. Figure 1 depicts a possible configuration of the offshore hub: an artificial island will be built to collect the wind power produced, while several *point-to-point* High-Voltage Direct-Current (HVDC) links will connect the island to the onshore grids [2], forming a zero-inertia offshore AC system [3].

Being a first-of-a-kind project, TSOs face a series of new technical challenges related to reliability of power systems with only converter-based resources. Indeed, in the event of any power outage (N-1), the system must be able to restore the power balance fast enough to preserve the transient stability of the system [4]. Due to the complete absence of synchronous generators, the system will need either additional devices, such as DC choppers or crowbars, or a coordinated control strategy for active power sharing among converters, such as slack bus or frequency droop control [5]. In case of slack bus control, one converter absorbs the complete deviation, while the others do not take part in the active power regulation. The main disadvantage of this approach is that the converter acting as slack bus must be significantly oversized to satisfy the N-1 security criterion. Another option to ensure the secure operation of the autonomous offshore AC grid, that is formed by multiple grid-forming converters, is to utilize a coordinated control design that enables the active and reactive power sharing among them [6]. To enable that, different control strategies can be used [7]; namely virtual impedance control, angle droop, washout filter-based method, frequency-droop together with virtual impedance control scheme and PD type frequency droop. In this paper, we select the frequency droop together with the virtual impedance control scheme, since it does not require communication between the converters and provides robustness to system parameters. Frequency droop control allows the converters to share the power imbalance according to their droop gains, mimicking the behavior of synchronous generators. As for now, it appears as the preferable option for maintaining the power balance in the system.

The selection of frequency droop gains is mainly determined based on small-signal stability and dynamic performance of the system [8]–[13]. In addition to this, other limitations must be taken into consideration when tuning the frequency droops of converters in zero-inertia systems. With the main com-

G. S. Misyris, A. Tosatto and S. Chatzivasileiadis are with the Technical University of Denmark, Department of Electrical Engineering, Kgs. Lyngby, Denmark (emails: {gmisy, antosat, spchatz}@elektro.dtu.dk). T. Weckesser is with Dansk Energi, Frederiksberg C, Denmark (email: twe@danskenergi.dk).

This work is supported by the multiDC project, funded by Innovation Fund Denmark, Grant Agreement No. 6154-00020B.

Submitted to "IEEE Transactions on Power Systems" on February 19, 2021, and revised on July 21, 2021.

ponents of power converters being semiconductor switches, current limits represent a strict constraint for operation. As described in [14], under large disturbances, such as a trip of a converter (i.e. overload incident), $Pf$ droop-controlled grid-forming converters may lose synchronism, which results in the violation of the N-1 criterion. This occurs when the converter reaches its current limits, i.e. saturates, which in turn leads to a different maximum active power transfer capability of the converter. As a consequence, this limits how much the angle can vary after the disturbance, before the grid-forming converter loses the synchronism. The authors in [14] conclude that the $Pf$ droop-controlled grid-forming converters can remain synchronized (stable) only if the angle can converge to an equilibrium.

To enhance the stability of grid-forming converters during overload incidents, researchers have focused on developing various current limiting control strategies. Current limitation strategies define the overload capability of the converters following a large-signal disturbance, which is primarily characterized by the output impedance characteristics of the grid-forming converter [15]. These characteristics are determined based on the network layout and the current limitation strategy employed by the grid-forming converter. As described in [16], there are two main strategies to limit the current following a large-signal disturbance: (i) switching the control structure of the grid-forming converter to grid-following mode [17] or (ii) limiting the current references of the grid-forming converter during the fault, i.e. the current of the grid-forming converter reaches a maximum value and is limited by the current limitation strategy [15], [18]–[21]. The latter strategy is adopted by most of the researchers, because it preserves the functionality of the grid-forming control even if the converter hits the current limits and saturates.

For the methods that directly limit the current references, several current limitation strategies have been proposed. In [15] and [18], the authors proposed the use of a dynamic virtual impedance which limits the current references of the grid-forming converter and improves the transient stability of the system. However, as mentioned in [16] the accuracy of this current limitation strategy depends on several unknown factors and is challenging when multiple grid-forming converters are connected to the system. Moreover, the dynamic virtual impedance can cause small-signal instability as presented in [15]. The current limiting control strategies that have been presented in [16], [19]–[21] focus on adjusting the power references such that wind-up in the outer loops is avoided and the transient stability of the system is enhanced. This is done by either adjusting the power references based on voltage and current measurements or by updating the droop gains in the outer control loop of the grid-forming converter. Although the schemes proposed by the authors improve the transient stability of grid-forming converters using only local measurements, they have only been evaluated in systems containing synchronous generators or an infinite bus, while the offshore system will consist of converter-based resources. Thus, this paper aims at investigating and improving the synchronization stability of zero-inertia offshore AC systems, that consist of multiple grid-forming converters operating in parallel. In particular, we focus on updating the frequency droops in the outer control loops of the grid-forming converters, while considering their available headroom for contributing to active power regulation and the dynamic performance of frequency droop control.

Moreover, the active power set-point of the HVDC converters will be determined by the market and will vary significantly based on the total wind power produced and the resulting exchanges between NSWPH-connected countries. The question that arises is: will it be possible to define an unique set of gains that never violates the operating limits of the converters in case of contingency? Given that TSOs will want to guarantee the N-1 security criterion without relying on current limiting control techniques, the only possibility in case of fixed gains is to increase the Transmission Reliability Margin (TRM) and decrease the capacity available for market operation. This could lead to wind curtailment and market inefficiencies, calling for a new centralized approach to update the frequency droop values and adjust the active power set-points in case of contingency, ensuring reliable system operation in the island.

Centralized control strategies have already been deployed by TSOs, an example is the MIO controller of the back-to-back converter installed in Kriegers Flak CGS [22]. In addition, the concept of adaptive droop settings has been previously proposed for multi-terminal DC systems [23]–[26], where the DC voltage droops are updated considering the available headroom of the converters. In this context, this paper combines both approaches and proposes a centralized method for updating the frequency droop values of HVDC converters in offshore systems, with the goal of distributing the active power in a way that does not violate the operational limits of the converters in case of contingency, and preserve the synchronization stability of the system. The calculation of frequency droop gains is performed solving an optimization problem which takes into consideration the available headroom of each converter and guarantees N-1 security. Dynamic and market simulations are performed to show the benefits of the proposed approach both in terms of dynamic response of the system and market operation. In detail, the contributions of this paper can be summarized as follows:

- a methodology to determine the frequency droop values and distribute any power imbalance without violating the operational limits of the remaining converters at the post-fault state.
- an analysis of the dynamic performance of the proposed methodology using the $H_2$ norm of the system as a performance metric.
- a cost benefit analysis which compares the proposed methodology to the common paradigm (static equal droops) in terms of market operation, highlighting the benefits of increased transmission capacity.

The rest of this paper is organized as follows. Section II introduces the operating principles of offshore and onshore converters. In section III, the optimization problem for selecting the droop gains of the converters is presented. The results of the dynamic and market simulations are discussed in Section IV. Section V concludes the paper.

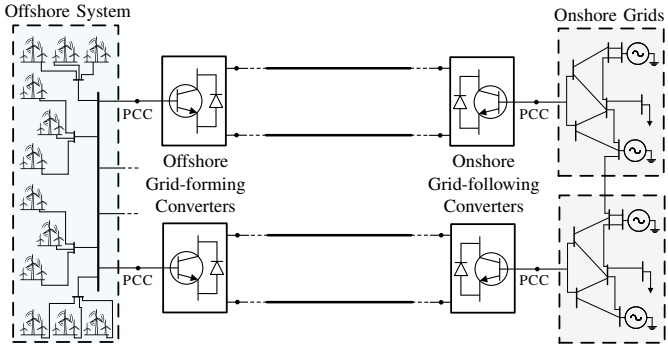

Fig. 2. Offshore energy islands with point-to-point HVDC connections to the onshore grids.

## II. OFFSHORE SYSTEMS - OPERATING PRINCIPLES

An offshore system consists of multiple offshore wind farms, a collection grid that transfers the produced power to a single or multiple offshore substations, and several HVDC links which transfer the collected wind power to the onshore systems, as depicted in Fig. 2. In the specific case of the NSWPH, the project consortium is considering to build one or multiple artificial islands instead of having floating substations. Multiple grid-forming converters are placed on these islands and are used to set the frequency of the offshore system. In addition, we consider that the wind farms consist of Type-IV wind turbines, where the power from the wind generator is passed through a back-to-back converter before being injected into the offshore grid. Considering that the mechanical part of the wind-turbine is decoupled from the offshore system and it is only interfaced through a voltage-source converter that operates in grid-following mode, the wind turbine will not inherently provide rotational inertia, unless it is controlled for that specific purpose [27]. In this paper, we consider that they do not exhibit an inertial response, thus, the offshore system is a zero-inertia AC system.

In zero-inertia systems, offshore converters share any power mismatch using the frequency droop control scheme: the active power exchanged with the AC system after a contingency depends on the frequency droop gain of each converter. However, when multiple grid-forming converters are connected to the same AC system, it becomes difficult to define the exact amount of power that each converter should handle after a contingency (wind farm loss, converter outage, etc.), which makes the system unreliable. Reliability of electricity supply is a fundamental requirement of power system operation, such that TSOs enforce different security criteria to limit the impact of disturbances. Among others, the N-1 criterion establishes that the system must be capable of withstanding the loss of a single component without violating operational security limits. Using fixed frequency droops might require additional control actions to be in compliance with the N-1 security criterion, such as blocking multiple offshore converters, or curtail wind in order not to exceed the current limits of the offshore converters. As a result, there is a need of adaptive frequency-droop based controller that accounts for the headroom of the converters and the varying profile of wind power generation.

Similarly to [23], in our studies, we consider a converter station outage as a critical contingency that the system should withstand (N-1 security).

In the following, the control principle of the converters connected to the offshore and onshore systems are presented. In this work, we consider that the wind farms connected to the hub operate at unity power factor ($Q = 0$) at pre-fault state and do not participate in the primary frequency control.

### A. Onshore VSC Control Scheme

VSCs connected to the onshore AC grids operate in grid-following mode [28]. In such operating mode, the main goal of the active power controller is to balance the voltage on the DC side. For point-to-point HVDC connections, PI controllers are deployed to control the DC voltage of the link and the reactive power exchanged with the onshore AC grid to their reference values, respectively.

### B. Offhore VSC Control Scheme

VSCs connected to the offshore AC network operate in grid-forming mode [28]. The basic control structure of grid-forming converter consists of an active power controller, a voltage controller and a current limiter controller that saturates the converter in case it exceeds its nominal current value. More details on the grid-forming control structure can be found in [10] and [19]. In case of power imbalance, the main objective of a grid-forming converter is to remain synchronized with the rest of the grid without violating its operational limits, i.e. the converter will not share the burden of the converter loss if it operates close to its limits. To ensure this, we propose an adaptive primary frequency droop-based controller that considers the dynamic performance of the system and complies with the N-1 security criterion in case of converter outage. By adapting the frequency droop values, we aim at limiting the angle variation of the converters with the least available current headroom at the pre-fault state. By using a fixed set of droops, instead, the angle variation does not depend on the loadability of the converters, i.e. it does not account for the available headroom of the offshore converters, thus the offshore converters are more likely to saturate, which can result in synchronization instability. By adapting the frequency droop values, we ensure that the converter current does not exceed the limits at the post-fault state, and remain synchronized with the grid [14]. A frequency droop gain is assigned to each converter, and the post-contingency steady-state power output of each converter will then depend on the combined effect of the droops of the grid-forming converters.

## III. PROPOSED METHODOLOGY

The scope of this work is to develop a methodology for calculating frequency droop gains of offshore converters in zero-inertia offshore grids which enable current distribution among the HVDC links without exceeding the converter limits, as well as enhanced dynamic performance during small disturbances. In this section, we present the optimization problem used to determine the set of droop gains with these properties.

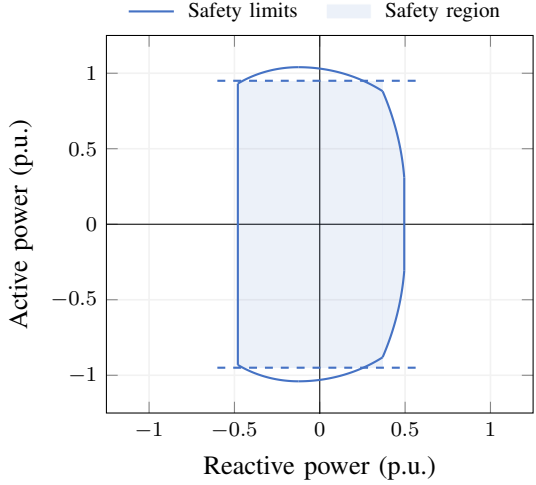

Fig. 3. PQ diagram of offshore VSCs [29].

## A. Frequency Control in Zero-Inertia Offshore Systems

Similar to synchronous machine-based systems, the frequencies imposed by grid forming VSCs must converge to a common value $\omega$ at steady-state. The relationship between the active power absorbed/injected by the $i$-*th* offshore converter, $P_i^*$, and the imposed frequency, $\omega_i$, is given by:

$$P_i^* = P_i^{\text{ref}} + \frac{1}{k_{f_i}} \Delta \omega_i \qquad (1)$$

where $k_{f_i}$ is the frequency droop gain, $P_i^{\text{ref}}$ is the reference value of the active power and $\Delta \omega_i = \omega^{\text{ref}} - \omega_i$, with $\omega^{\text{ref}}$ the chosen reference frequency. At steady state $\Delta \omega_i = \Delta \omega$, with $\Delta \omega$ being the deviation of the average frequency of the offshore system.

To preserve the stability of the offshore system, the sum of the active power injected/withdrawn by the converters must be equal to the wind power produced. In case of converter ($k$) outage, the mismatch between the power produced by the wind farms and the power absorbed by the converters results in a frequency deviation equal to:

$$\Delta \omega = \frac{1}{\sum_{i \neq k} \frac{1}{k_{f_i}}} P_k^{\text{ref}}. \qquad (2)$$

where $P_k^{\text{ref}}$ is the reference set-point of the $k$-*th* converter at pre-fault state. In case of converter outage, thus, the steady-state frequency deviation is inversely proportional to the sum of the inverse droop gains of the remaining converters connected to the offshore grid. It follows that the post-fault power flowing through the $i$-*th* converter, $P_i^*$, is given by:

$$P_i^* = P_i^{\text{ref}} + \frac{1}{k_{f_i}} \frac{1}{\sum_{m \neq k} \frac{1}{k_{f_m}}} P_k^{\text{ref}}. \qquad (3)$$

In the following, the proposed methodology for calculating the droop gains of offshore converters is presented.

## B. Optimization Problem

The selection of the frequency droop gains of converters plays a key role in ensuring small-signal stability and reliable operation of zero-inertia offshore systems [17].

The safe operating region of VSC converters is depicted in Fig. 3. On the y-axis, the active power output is limited by the maximum current the power electronic components can withstand: as the heating is proportional to the square of the current magnitude, the active power output slightly decrease with increasing reactive power output. On the x-axis, reactive power limitations are driven by voltage constraints. To ensure that the converter can contribute to voltage control by adjusting its reactive power, the active power limit is set to 0.95 pu.

On the one hand, too small values of frequency droop gains can lead to operating the converter outside its safety region; on the other hand, too high values of frequency droops make the system small-signal unstable due to the high influence of eigenfrequencies on the angle dynamics of the converter. This has already been illustrated in numerous works [8]–[10], [19], [30], [31] and, thus, the analysis for identifying the stability margins is omitted in this paper. To allow for larger values of the frequency droops, we can use an active damping controller, such as the static virtual impedance controller proposed in [15] and [32]. This results in enhancing the small-signal stability of the system, i.e. achieving a sufficient level of damping to avoid power oscillations. We empirically consider a maximum value of the frequency droop equal to 0.1, similar to [4] and [31].

The selection of adequate droop gains can be performed by solving an optimization problem where the conditions to ensure small-signal stability of the system and N-1 security are enforced as constraints. Such an optimization problem, where frequency droop gains (more precisely, their reciprocal $x$) are the decision variables, can be formulated as:

$$\min_x \sum_{i=1}^{n-1} \sum_{c=i+1}^{n} |x_i - x_c| \qquad (4a)$$

$$\text{s.t.} \quad \alpha = \sum_i x_i, \qquad (4b)$$

$$x_i \geq \underline{X}_i, \qquad : \forall i \qquad (4c)$$

$$\left| P_i + \frac{x_i}{\alpha - x_k} P_k \right| \leq \overline{P}_i^{\max}. \qquad : \forall i, \forall k \qquad (4d)$$

For simplicity, droop coefficients $k_f$ have been replaced by their reciprocal $x = 1/k_f$.

The objective function (4a) represents the sum of the distances between the droop gains. The goal of the optimization problem is, thus, to find a feasible set of droop gains close to "equal droop gains". This is done in order to have an equal distribution of the post-fault extra power in case constraints (4d) are not binding. As it will be shown later in section III-C, this control objective helps avoid power oscillations between the offshore converters and improve the dynamic performance of the frequency droop controllers [33].

The maximum Steady-State Frequency Deviation (SSFD) after the occurrence of an active power disturbance in the offshore network is a function of the sum of the droop gains ($\alpha$), which is also a measure of the system stiffness. Thus, constraint (4b) makes sure that the selected droop gains sum up to the desired value (in this work, $\alpha$ is considered equal to 600, which is equivalent to a maximum steady state deviation of 0.0833 Hz after a step change of 2000 MW). Constraints (4c), instead, ensure small-signal stability, representing the upper bound on the droop gains ($\underline{X}_i$ is the lower bound on $x_i$, the reciprocal of the droop gain $k_{f,i}$). Finally, constraints (4d) are derived from equation (3) and limit the selection of the droop gains to those value which guarantee safe post-fault operation.

Problem (4) is non-linear, as constraints (4d) contain variables both in the numerator and denominator. By defining:

$$\alpha_k = \alpha - x_k, \tag{5a}$$
$$\sigma_k = \frac{1}{\alpha_k}, \tag{5b}$$
$$z_{k,i} = \sigma_k x_i, \tag{5c}$$

constraint (4d) can be rewritten as:

$$|P_i + z_{k,i} P_k| \leq \overline{P}_i^{\max}. \quad : \forall i, \forall k \tag{6}$$

Although Eq. (6) is now linear, Eq. (5b) and (5c) are bilinear. In order to linearize them, the multi-parametric disaggregation technique presented in [34] is used. Eq. (5b) and (5c) are thus recast into:

$$1 = \sum_b \sum_a \hat{\sigma}^\alpha_{k,a,b} \cdot a \cdot 10^b, \tag{7a}$$
$$z_{k,i} = \sum_d \sum_a \hat{\sigma}^x_{k,i,a,d} \cdot a \cdot 10^d, \tag{7b}$$

In order to keep consistency with Eq. (5b), the following set of equalities and inequalities should be included together with Eq. (7a):

$$\alpha_k = \sum_b \sum_a a \cdot 10^b \cdot y^\alpha_{k,a,b}, \tag{8a}$$
$$\sigma_k = \sum_a \hat{\sigma}^\alpha_{k,a,b}, \quad : \forall b \tag{8b}$$
$$0 \leq \hat{\sigma}^\alpha_{k,a,b} \leq \overline{S}_k \cdot y^\alpha_{k,a,b} \quad : \forall a, \forall b \tag{8c}$$
$$\sum_b y^\alpha_{k,a,b} = 1, \quad : \forall b \tag{8d}$$
$$y^\alpha_{k,a,b} \in \{0,1\}. \quad : \forall a, \forall b \tag{8e}$$

with $a \in \{0,1,...,9\}$ and $b \in \{\psi_b, ..., \eta_b\}$. In Eq. (8a), $\alpha_k$ is expressed as a multi-parametric sum of active decimal powers determined by the binary variables $y^\alpha_{k,a,b}$, while $\sigma_k$ is disaggregated into a set of continuous non-negative variables, represented by $\hat{\sigma}^\alpha_{k,a,b}$, in Eq. (8b). The two parameters $\psi_b$ and $\eta_b$ denotes the powers of ten used for the parameterization of $\alpha_k$. Eq. (8c) enforces the limits on $\hat{\sigma}^\alpha_{k,a,b}$, which must be non-negative and equal to the upper bound $\overline{S}_k$ at most. From Eq. (5b), $\overline{S}_k = \frac{1}{\sum_{i \neq k} \underline{X}_i}$. Finally, Eq. (8d) guarantees that only one binary variable is active for the $b$-th place in the power representation of $\hat{\sigma}^\alpha_{k,a,b}$.

Similarly, together with Eq. (7b) the following equalities and inequalities are included:

$$x_i = \sum_d \sum_a a \cdot 10^d \cdot y^x_{i,a,d}, \tag{9a}$$
$$\sigma_k = \sum_a \hat{\sigma}^x_{k,i,a,d}, \quad : \forall d \tag{9b}$$
$$0 \leq \hat{\sigma}^x_{k,i,a,d} \leq \overline{S}_k \cdot y^x_{i,a,d} \quad : \forall a, \forall d \tag{9c}$$
$$\sum_d y^x_{i,a,d} = 1, \quad : \forall d \tag{9d}$$
$$y^x_{i,a,d} \in \{0,1\}. \quad : \forall a, \forall d \tag{9e}$$

with $a \in \{0,1,...,9\}$ and $d \in \{\psi_d, ..., \eta_d\}$. The two parameters $\psi_b$ and $\psi_d$ are chosen according to the desired decimal precision, e.g. $\psi_b = \psi_d = -5$ denote a precision of $10^{-5}$, while $\eta_b$ and $\eta_d$ are determined by the magnitude of $\alpha_k$ and $x_i$, respectively. For example, the upper bound of $\alpha_k$ is $\alpha - \underline{X}_k$; if $\alpha = 300$, $\eta_b$ must be equal to or greater than 2.

After the linearization, problem (4) is recast into:

$$\min_\Gamma \sum_{i=1}^{n-1} \sum_{c=i+1}^{n} |x_i - x_c| \tag{10a}$$
$$\text{s.t. } (4b) - (4c) \tag{10b}$$
$$\quad (5a) \quad : \forall k \tag{10c}$$
$$\quad (7a), (8a)-(8e) \quad : \forall k \tag{10d}$$
$$\quad (7b), (9a)-(9e) \quad : \forall i, \forall k \tag{10e}$$
$$\quad (6) \tag{10f}$$

with $\Gamma = \{x, \hat{\sigma}^\alpha, \hat{\sigma}^x, y^\alpha, y^x\}$. Problem (10) is now a Mixed Integer Linear Problem (MILP) and can be solved with commercial and open-source MILP solvers (e.g. Gurobi, Mosek or GLPK).

### C. Dynamic Performance of Frequency Droop Control

As explained in section III-B, the objective of the optimization problem is to minimize the distance between the droop gains of the converters. To motivate this choice, we now evaluate the performance of primary frequency control when offshore converters have equal or different frequency droop values, using as a metric the $H_2$ norm of the system, i.e. root-mean-square of the impulse response of the system.

The simplified model used in this section, depicted in Fig. 4, considers an AC system with $n$ nodes "formed" by grid-forming converters. Similar to [35] (that also considers an AC network with multiple grid-forming converters and loads), the AC system is described by linear power flow and Kron-reduced network models. Several wind farms are connected to the system, modelled as negative constant loads. Under the assumptions of linear power flow model (i.e. voltage magnitude are considered constant and cable capacitance negligible), any constant power load can be represented by a constant impedance load and, under the assumption of Kron-reduced network model, included in the impedances of the cables. Since the frequency is regulated by the grid-forming converters, the frequency control dynamics can be described as:

$$\dot{\theta}_i = \omega_i \tag{11}$$
$$T_{p_i} \dot{\omega}_i = -\omega_i + \omega_{\text{ref}} + k_{f_i}(P_i^{\text{ref}} - P_i) \tag{12}$$

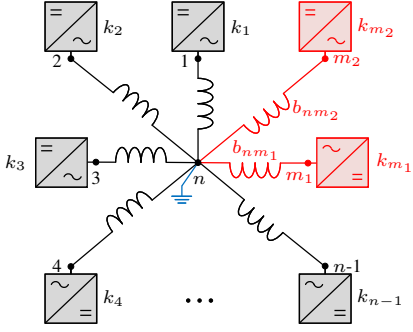

Fig. 4. Example of a network of $n$ converter nodes. Node $n$ is grounded following the approach in [36]. Two additional converter nodes $m_1$ and $m_2$, highlighted in red, are included in the system to study its dynamic performance with equal or different droop gains.

where $T_{p_i}$ represents the delay for measuring the active power $P_i$ at the PCC of the grid-forming converter. Based on the linear power flow assumptions, $P_i = \sum_j b_{ij}(\theta_i - \theta_j)$, where $b_{ij}$ is the susceptance of the cable connecting node $i$ to node $j$ and $\theta_i, \theta_j$ are the voltage angles at node $i$ and $j$.

For the dynamic performance analysis, it is assumed that the system is subject to persistent amplitude noise (wind fluctuation), modelled as an additive disturbance $w(t)$. Similar to [33], the system output $y(t)$ is the frequency deviation $\omega(t)$ at the terminal of grid forming converters, since the focus of the analysis is on the impact of the frequency droop gains on the frequency fluctuations and, thus, active power oscillations.

Considering a second-order frequency droop control model for inverter-based networks [35], the state-space model of the system model can be described as follows:

$$\begin{bmatrix}\dot{\theta}\\\dot{\omega}\end{bmatrix} = \underbrace{\begin{bmatrix}0 & I\\-T_p^{-1}K_f L_B & -T_p^{-1}\end{bmatrix}}_{A}\begin{bmatrix}\theta\\\omega\end{bmatrix} + \underbrace{\begin{bmatrix}0\\-T_p^{-1}K_f\end{bmatrix}}_{B}w \quad (13a)$$

$$y = \underbrace{\begin{bmatrix}0 & I\end{bmatrix}}_{C}\begin{bmatrix}\theta\\\omega\end{bmatrix} \quad (13b)$$

where $T_p = \text{diag}(T_{p_i})$, $K_f = \text{diag}(k_{f_i})$ and $L_B$ is the $b_{ij}$-weighted Laplacian matrix. We assume that all the time delays are equal, i.e. $T_{p_i} = \tau\ \forall i$, and that the grid-forming converters do not provide an inertia effect. To evaluate the dynamic performance of the system with a certain set of droop gains, the $H_2$ norm of the system is computed. The $H_2$ norm, which gives an interpretation of how the gains affect the frequency and active power response after an active power disturbance, is calculated as follows:

$$\|H\|^2_{\mathcal{H}_2} = \text{tr}(B^T X B) \quad (14)$$

where $X$ is the observability Gramian computed by solving the Lyapunov equation:

$$A^T X + X A = -C^T C \quad (15)$$

*1) Performance analysis:* We first look at how the dynamic performance of the system is affected by different parameters, such as the number of grid-forming converters, the value of the droop gains and the inertia effect of the converters. Following the approach in [36], we use a reduced model of the system described by (13). The reduced order model is based on the change of variables:

$$\theta = U\theta' \quad \text{and} \quad \omega = U\omega' \quad (16)$$

where $U$ is the unitary matrix that diagonalizes $L_B$. This change aims at eliminating the zero eigenvalue contained in the matrix $A$. The physical interpretation of the reduced system is that one of the nodes in the network is considered grounded.

Assuming that all the droop gains are equal, i.e. $k_{f_i} = k\ \forall i$, the state-space model describing the reduced model is:

$$\begin{bmatrix}\dot{\theta}'\\\dot{\omega}'\end{bmatrix} = \begin{bmatrix}0 & I\\-\frac{k}{\tau}\widetilde{L}_B & -\frac{1}{\tau}I\end{bmatrix}\begin{bmatrix}\theta'\\\omega'\end{bmatrix} + \begin{bmatrix}0\\-\frac{k}{\tau}I\end{bmatrix}w' \quad (17a)$$

$$y' = \begin{bmatrix}0 & I\end{bmatrix}\begin{bmatrix}\theta'\\\omega'\end{bmatrix} \quad (17b)$$

From (14), the $H_2$ norm of the system is calculated as:

$$\|\widetilde{H}\|^2_{\mathcal{H}_2} = (n-1)\frac{(k)^2}{2\tau} \quad (18)$$

The following observations can be made for the $H_2$ norm: (i) it is directly proportional to the value of the frequency droop gain $k$, (ii) it decreases with increasing time delay $\tau$ and (iii) it increases with increasing number of grid-forming converters. The first observation is related to the responsiveness of the controller: the higher the frequency droop gain, the faster the response of the converter and, thus, the higher the oscillations in the frequency. The second observation has to do with the inertia effect provided by the grid-forming converters: the higher the time delay, the higher the inertia effect and, thus, the smaller the frequency fluctuations. Finally, the third observation suggests that the droop gains should be adjusted in case a grid-forming converter is added/removed from the system to preserve the performance of the frequency droop control strategy.

*2) Comparison between equal and different gains:* To study how the dynamic performance of the system is affected by selecting different droop gains for the converters, we consider two additional nodes $m_1 = n+1$ and $m_2 = n+2$ connected to the grounded node (for simplicity node $n$). Two grid-forming converters, with droop gains $k_{m_1}$ and $k_{m_2}$ respectively, are connected to these nodes. Similar to the other nodes, also the new converters are subject to persistent amplitude noise, respectively $w_{m_1}$ and $w_{m_2}$. The new system $\widehat{H}$ consists of system $\widetilde{H}$ described by (17) and two subsystems formed by the two grid-forming converters, respectively $H_{m_1}$ and $H_{m_2}$. The state-space model of these subsystems can be written as:

$$\begin{bmatrix}\dot{\theta}'_m\\\dot{\omega}'_m\end{bmatrix} = \begin{bmatrix}0 & 1\\-\frac{k_m}{\tau}b_{nm} & -\frac{1}{\tau}\end{bmatrix}\begin{bmatrix}\theta'_m\\\omega'_m\end{bmatrix} + \begin{bmatrix}0\\-\frac{k_m}{\tau}\end{bmatrix}w'_m \quad (19a)$$

$$y'_m = \begin{bmatrix}0 & 1\end{bmatrix}\begin{bmatrix}\theta'_m\\\omega'_m\end{bmatrix} \quad (19b)$$

where $m \in \{m_1, m_2\}$. As described in [36], system $\widehat{H}$ can be decoupled into three subsystems, $\widehat{H} = \text{diag}(\widetilde{H}, H_{m_1}, H_{m_2})$. It follows that the squared $H_2$ norm of system $\widehat{H}$ can be

computed as:
$$||\widehat{H}||^2_{\mathcal{H}_2} = ||\widetilde{H}||^2_{\mathcal{H}_2} + ||H_{m_1}||^2_{\mathcal{H}_2} + ||H_{m_2}||^2_{\mathcal{H}_2}. \quad (20)$$

Similar to (18), the sum of the squared norms of subsystems $H_{m_1}$ and $H_{m_2}$ can be calculated as follows:
$$||H_{m_1}||^2_{\mathcal{H}_2} + ||H_{m_2}||^2_{\mathcal{H}_2} = \frac{(k_{m_1})^2}{2\tau} + \frac{(k_{m_2})^2}{2\tau} = \frac{\varepsilon}{2\tau} \quad (21)$$
where $\varepsilon$ is equal to the sum of the squared frequency droops values $k_{m_1}$ and $k_{m_2}$.

Let us now assume that the frequency droop gains sum to a fixed value $\sigma$, which corresponds to fixing the SSFD to a specific value. If $k_{m_1} > k_{m_2}$, it follows that $||H_{m_1}||^2_{\mathcal{H}_2} > ||H_{m_2}||^2_{\mathcal{H}_2}$, which indicates that the performance of the frequency controller of converter $m_1$ is worse compared to the one of $m_2$. Moreover, the relation between $\varepsilon$ and $\sigma$ is given by:
$$\varepsilon = \sigma^2 - 2k_{m_1}k_{m_2}. \quad (22)$$

Equation (22) clearly shows that the minimum value of the sum of the squared $H_2$ norm of the two subsystems corresponds to the situation where $k_{m_1} = k_{m_2}$, meaning that the dynamic performance of the system is improved when all the converters have the same droop gain.

### D. Timeline for the calculation of the frequency droops

The timeline for the calculation of the frequency droop coefficients is depicted in Fig. 5. The calculation of the droop coefficients is performed together with the calculation of the available transmission capacity for exchanges before the day-ahead market is cleared. The input necessary to solve this optimization problem is a set of possible trades on the HVDC links, i.e. a forecast of the exchanges in the day-ahead market determined based on demand and wind scenarios. The optimization problem is then solved considering: (i) the dynamic performance of the controller (see Eq. (4a)), (ii) the steady-state frequency deviation (see Eq. (4b)), (iii) the system small-signal stability (see Eqs. (4c)) and (iv) the available headroom of the converters at pre-fault state (see Eqs. (4d)). If the optimization problem is infeasible, it means that there is not enough capacity for contributing to the active power regulation in the network and, therefore, the transmission capacity is reduced and the HVDC set-points are forecasted again. This process is repeated until the optimization problem gives a feasible solution. The available transmission capacity is then communicated to the market operator for the day-ahead market clearing, which determines the exchanges on the HVDC links.

While solving the optimization problem, the solver will try to minimize the distance between the frequency droop coefficients, meaning that all HVDC converters will have the same droop coefficient if their operational limits are not violated. Otherwise, the converter with the highest headroom will receive the smallest droop coefficient, while the most loaded converter will have the largest one (reducing its contribution at post-fault). When the flows are determined by the market operator (after the day-ahead market has been cleared), the optimization problem could be solved again with the aim of further reducing the distance between the droop coefficients. Given that the set-points of HVDC links could be further modified by the exchanges in the intra-day market, the droop coefficients could be updated close to real-time operation.

Adjusting the droop coefficients every time the HVDC set-points are modified maximizes the utilization of the interconnectors, as this can significantly mitigate the limitations that security constraints impose on market operation. Considering that the frequency droops are calculated prior to the occurrence of the contingency (before real-time operation), the response speed of the active power controller of the grid-forming converter remains equally fast and is not affected at all by how fast the optimization problem is solved.

## IV. SIMULATION AND RESULTS

The test system used for the validation of the proposed methodology is inspired by the NSWPH project, and represents an offshore energy island configuration for massive integration of offshore wind power. Ten wind farms, with a total installed capacity of 9 GW, are connected to the island through 400 kV HVAC cables. Six point-to-point HVDC links connect the island to the onshore grids, as depicted in Fig. 6. The rated power of each offshore and onshore converter is 1850 MVA, and the base power of the system is $S_b$ = 1850 MVA. More details about the converters' control structure and the models of power system components can be found in [4]. All the dynamics simulations presented in this section have been performed with DIgSILENT PowerFactory [37].

A market model representing the European electricity market in 2030 is used to determine the power flows from the island to each country for a period corresponding to one year. In order to highlight the problems arising with static equal droops, the market is first cleared without considering the N-1 security criterion, i.e. without limitations on the available transmission capacity. Subsequently, the compliance with the N-1 criterion is checked before clearing the market and transmission capacities are decreased if necessary. The market is then cleared again to highlight the benefit of the proposed methodology. An example of the power flows during one day (24 hours) is plotted in Fig. 7. From this representative day, the hour with the most critical set of flows has been selected for the dynamic simulations, which corresponds to limitations, the legend is omitted in the next plots; the reader is referred to the legend of Fig. 7.

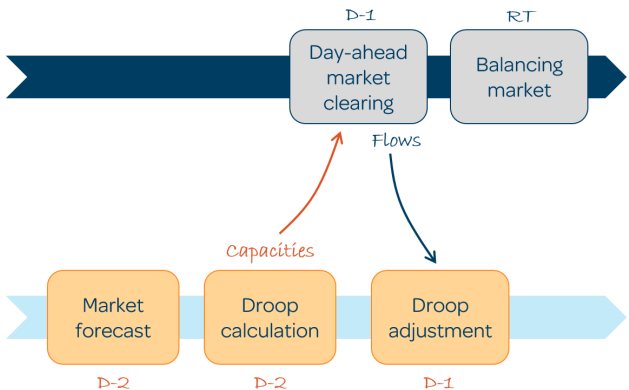

Fig. 5. Timeline for the calculation of the frequency droop coefficients.

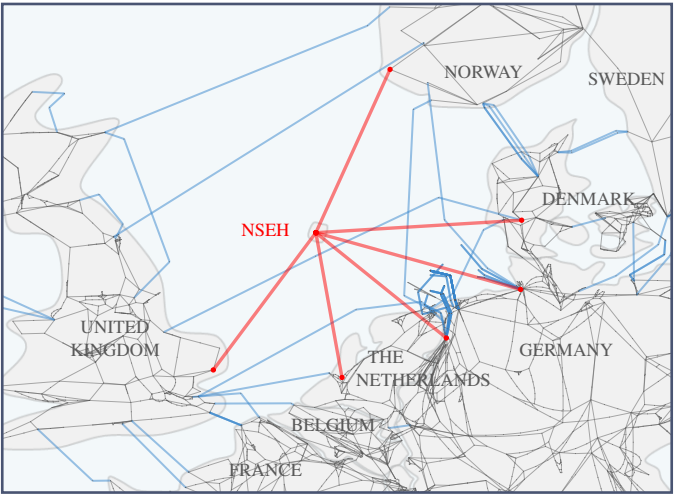

Fig. 6. Connections between the offshore island and the onshore grids. The electrical networks correspond to the 400 kV transmission networks of the respective countries.

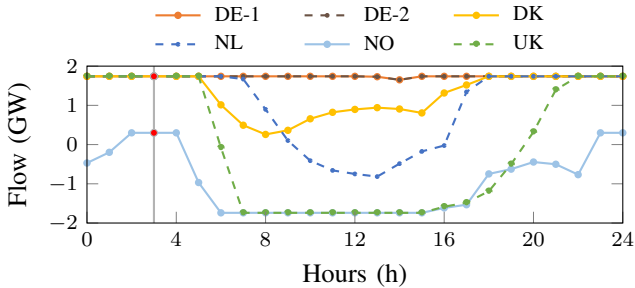

Fig. 7. Flows over the HVDC interconnectors.

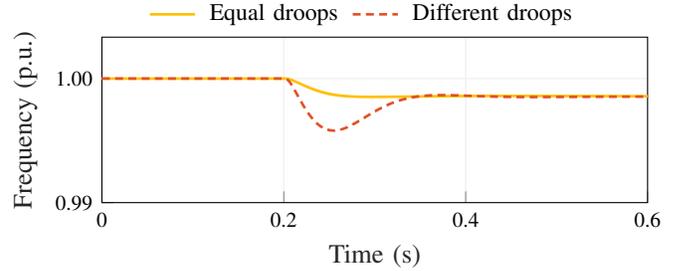

Fig. 8. Average frequency deviation following a step decrease of wind power in case of equal and different droops.

### A. Dynamic Performance of frequency droop control

In the first simulation, a step decrease (250 MW) of the wind power output is applied to the system. Fig. 8 shows the frequency deviations of the grid-forming converters. The system stiffness is the same in both simulations ($\alpha = 600$); however, in the first case all the droop coefficients are equal, while in the second case they are different. As expected, the average steady-state frequency deviation in the system is the same in both cases. This confirms that the SSFD only depends on the system stiffness and it is not affected by individual droop values. Moreover, it is evident that the average frequency deviation of the system is higher in case of different gains. This confirms that the dynamic performance of the system is worse when grid-forming converters have different gains, as described in Section III-C2.

### B. Validation of the Proposed Methodology

In the second simulation, the effectiveness of the proposed frequency controller with adaptive droop gains is demonstrated. Two cases are analyzed: (i) frequency droop values are kept fixed over time and are all equal and (ii) frequency droops are updated based on the optimization problem presented in Section III every time the HVDC set-points are changed. Moreover, two different current limiting controllers, presented respectively in [14] and [15], are implemented for each case. This is done in order to show that the presented results are independent on the implemented current limiting strategy.

Fig. 9 shows the system response to the trip of the converter connected to UK (1740 MW); for a better visualization, only the converters connected to NL and NO are displayed. During the first milliseconds after the outage, the power that was previously flowing to UK is redirected to the other converters. In the case of equal frequency droops, it can be seen that during the first few hundred milliseconds the converters are saturated to 1.1 pu, which is the maximum saturation current. However, we can see that afterwards the converters lose synchronous stability with both the current limiting controllers. More details about the instability mechanisms in case of the dynamic virtual impedance current limiting controller and d-axis priority based current saturation algorithm, can be found in [19] and [14], respectively.

hour 3 when most of the converters are operated close to their limits. All the market simulations presented in this section have been performed using YALMIP [38] and Gurobi [39].

The remainder of this section is divided into three parts. In Section IV-A, we compare the dynamic performance of the frequency droop controllers with equal and different frequency droop gains in case of a sudden variation of the wind and different values of system stiffness. This is to highlight that equal droop gains better perform in terms of dynamic response. In Section IV-B, we evaluate the performance of the proposed methodology, considering a converter outage and the corresponding system response with fixed (equal) and adaptive gains. Finally, in Section IV-C, we assess the benefits of the proposed methodology in terms of increased transmission capability for market operation.

Without loss of generality, in all the simulations in Section IV-B and IV-C, we consider that the offshore converter at outage corresponds to the HVDC-link connecting the NSWPH to UK, with an active power output of 1740 MW. Thus, the loss of this converter is equivalent to an increase of active power (1740 MW) that must be shared among the other converters. It should be mentioned that, at the pre-fault state, all the converters control the voltage at their PCC, which is equal to 1 pu. This leads to a reactive power absorption in the range of 0-150 MVAr for all converters, which corresponds to an initial reactive current less than 0.08 pu. Finally, due to space

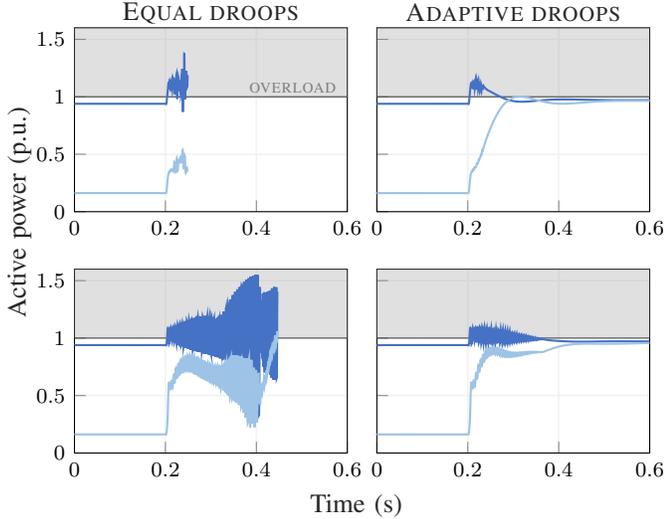

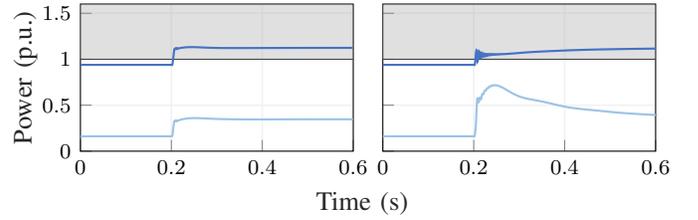

Fig. 10. System response to an offshore converter outage using the current limiting technique in [14] (left) and [15] (right). Both figures shows the power set-point of two of the remaining converters (NL and NO) with equal droops and $I_{max} = 1.2$ pu.

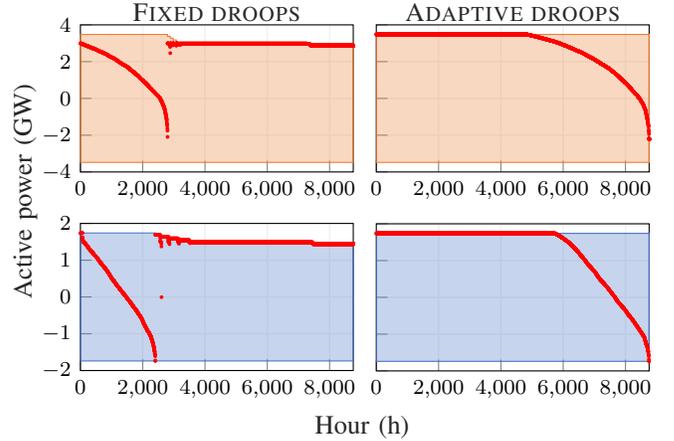

Fig. 9. System response to an offshore converter outage. The two left figures show the power set-point of two remaining converters (NL and NO) with equal droops, while the two right figures are obtained with adaptive droops. In both cases, the two current limiting techniques in [14] (upper figures) and [15] (lower figures) are used with $I_{max} = 1.1$ pu. The overload region is represented with the gray area.

In case of adaptive frequency droops, the system response initially triggers the applied current limiting schemes. However, due to the different gains, a big portion of the active power initially absorbed by the converter at outage is directed towards the offshore converter which had the largest headroom at pre-fault state. That "relieves" the offshore converters that were saturated immediately after the outage. As shown in Fig. 9, with the proposed frequency droop controller the active power absorbed by the remaining converters stays below its nominal value also at post-fault state regardless of what current limiting technique is implemented. This demonstrates that the effectiveness of the proposed method is not dependent on the current limiting strategy and that the power imbalance is distributed according to the available power headroom, ensuring N-1 security at every time interval.

### C. Impact of the Proposed Methodology on Market Operation

In the third simulation, the value at which the current saturates, $I_{max}$, is increased from 1.1 pu to 1.2 pu. This is done to show that, in case of contingency, the system remains stable if converters do not saturate. Fig. 10 shows the system response in case of equal droops with the two current limiting techniques. During the first milliseconds after the outage, all the offshore converters share equally the current of the converter at outage, which is distributed according to the frequency droop gains. Given that most of the converters were operating close to their limits (except NO), the active power limits of most of the converters are violated (we only represent NL as the other converters have the exact same behavior). Since now $I_{max}$ is increased, all the converters remain synchronised, which shows that the system is stable when the converters are not saturated. However, exceeding the active power limit of the converters violates the N-1 criterion. Thus, in case of equal droops, system operators would have to decrease the available transmission capacity in order to ensure that the converters will operate within the operating region after the contingency.

Fig. 11. Available transmission capacity (shaded area) and corresponding flows (red dots) to DE (upper plots) and NL (lower plots). In the left graphs, the available capacity has been calculated based on fixed/equal droop gains while, in the right ones, based on adaptive droop gains.

In order to understand what are the implications of reducing the available transmission capacity given to the market, the market simulations are run again now considering the N-1 criterion. In other words, the market is cleared and then the obtained flows are used to check the compliance with the N-1 security criterion. In both situations, with fixed equal droops and adaptive droops, the capacity is decreased (50 MW steps) on those converters whose power limits are violated in case of converter outage, and the market is cleared again with the new capacities. This is repeated until the N-1 criterion is fulfilled. This procedure could be intended as the offline calculation performed by TSOs to ensure the stability of the system before communicating the available transmission capacity to the market operator. Fig. 11 shows the available transmission capacities (shaded areas) between DE and NSWPH (upper graphs) and NL and NSWPH (lower graphs) and the resulting flows (red dots). These graphs show the duration curve of the capacities, meaning the values of the capacity have been rearranged in descending order. With fixed equal droops, the N-1 security criterion is often violated and the full capacity is allowed only for less than 30% of the time. It is interesting to observe that the full capacity is only allowed when it is not actually needed, as shown by the flows which are always below the maximum value. On the contrary, with adaptive droops, the gains are calculated in order to take advantage of the different headrooms, and the N-1 criterion is always satisfied. The reduction of transmission capacity in case of fixed droops results in more than 350 GWh of curtailed wind energy which, in turn, increase the total system costs by 85 million Euros. This could be avoided either by oversizing the converters, increasing the investment costs, or by adjusting the droop gains with the proposed methodology.

## V. Conclusion

In this paper, challenges with respect to active power sharing in zero-inertia systems have been discussed and a method for deriving and updating frequency droop gains has been developed, ensuring active power sharing in the system after a large disturbance in compliance with the N-1 security criterion.

The simulation results show that, in the event of an offshore converter outage, current limiting strategies lead to instability in case the converters are saturated. Thus, the N-1 criterion cannot be guaranteed. By considering the headroom of each converter when calculating the frequency droop gains, the proposed approach distributes the power among the converters while taking into consideration their safe operating regions. As a consequence, the proposed methodology avoids converter saturation and preserves their synchronization to the rest of the grid. Moreover, by enforcing upper bounds on the frequency droop gains, also the small-signal stability of the system is guaranteed. As a result, the proposed methodology allows for greater loadability of the offshore converters at pre-fault state and guarantees their safe operation in the event of any power imbalance. Finally, from an operational point of view, the additional capacity given to the market results in a better utilization of the interconnectors, avoids wind curtailment and reduces the total system costs without oversizing of the converters.